\documentclass[twocolumn]{aastex631}
\PassOptionsToPackage{table}{xcolor}
\usepackage[table]{xcolor}
\usepackage{tabularx,colortbl}
\usepackage{graphicx}
\usepackage{color,colortbl}
\usepackage{amsmath}
\usepackage{multirow}

\usepackage{enumitem}
\usepackage{graphicx}
\usepackage{dcolumn}
\usepackage{bm}
\usepackage{hyperref}
\usepackage{enumitem}

\usepackage{graphicx}
\usepackage{color}
\usepackage{amsmath}
\usepackage{multirow}
\usepackage{longtable}

\begin{document}
\title{Gravitational wave source populations: Disentangling an AGN component}

\author{V. Gayathri}
\affiliation{Department of Physics, University of Florida, PO Box 118440, Gainesville, FL 32611-8440, USA}
\affiliation{Leonard E. Parker Center for Gravitation, Cosmology, and Astrophysics, University of Wisconsin–Milwaukee, Milwaukee, WI 53201, USA}
\thanks{gayathri.v@ligo.org}
\author{Daniel Wysocki}
\affiliation{Leonard E. Parker Center for Gravitation, Cosmology, and Astrophysics, University of Wisconsin–Milwaukee, Milwaukee, WI 53201, USA}
\author{Y. Yang}
\affiliation{Department of Physics, University of Florida, PO Box 118440, Gainesville, FL 32611-8440, USA}
\author{R. O'Shaughnessy}
\affiliation{Center for Computational Relativity and Gravitation, Rochester Institute of Technology, Rochester, NY 14623, USA}
\author{Z. Haiman}
\affiliation{Department of Astronomy, Columbia University, 550 W. 120th St., New York, NY, 10027, USA}
\author{H. Tagawa}
\affiliation{Department of Astronomy, Columbia University, 550 W. 120th St., New York, NY, 10027, USA}
\author{I. Bartos}
\affiliation{Department of Physics, University of Florida, PO Box 118440, Gainesville, FL 32611-8440, USA}

\begin{abstract}
    The astrophysical origin of the over 90 compact binary mergers discovered by the LIGO and Virgo gravitational wave observatories is an open question. While the unusual mass and spin of some of the discovered objects constrain progenitor scenarios, the observed mergers are consistent with multiple interpretations.  A promising approach to solve this question is to consider the observed distributions of binary properties and compare them to expectations from different origin scenarios. Here we describe a new hierarchical population analysis framework to assess the relative contribution of different formation channels simultaneously. For this study we considered binary formation in AGN disks along with phenomenological models, but the same framework can be extended to other models. We find that high-mass and high-mass-ratio binaries appear more likely to have an AGN origin compared to the same origin as lower-mass events. Future observations of high-mass black hole mergers could further disentangle the AGN component from other channels.
\end{abstract}

\section{Introduction}
Understanding the origin of binary black hole mergers is the first step in utilizing black hole mergers to probe a range of astrophysical processes. The LIGO (\citealt{TheLIGOScientific:2014jea}) and Virgo (\citealt{TheVirgo:2014hva}) gravitational wave observatories have discovered about $90$ binary mergers so far (\citealt{LIGOScientific:2021djp}), providing important information on the astrophysical population of mergers. Binary black hole systems can form through various channels, including isolated stellar binaries (\citealt{1998A&A...332..173P,2002ApJ...572..407B,2016A&A...588A..50M,2016MNRAS.460.3545D}) or triples (\citealt{2014ApJ...781...45A,2016MNRAS.463.2443K,2020MNRAS.498L..46V}),  dynamical interactions in star clusters (\citealt{1993Natur.364..423S,2000ApJ...528L..17P}), primordial black holes formed in the early universe (\citealt{10.1093/mnras/168.2.399}), and in the accretion disks of active galactic nuclei (AGNs; \citealt{2012MNRAS.425..460M,2017ApJ...835..165B,2017MNRAS.464..946S,2020ApJ...898...25T,2020MNRAS.498.4088M,2022MNRAS.514.3886M}).

The increased number of binary black hole observations allows for a more detailed investigation of the population's mass and spin distributions. While individual events may provide anecdotal suggestions hinting at one formation channel or another, only an interpretation of the full census can enable one to disentangle the potential contributions from multiple formation scenarios. Several studies have previously explored how to disentangle multiple channels, largely relying on comparison to phenomenologically-motivated estimates of the detailed outcomes of full formation scenarios (\citealt{2020ApJ...893...35D,2021NatAs...5..749G,Gayathri:2021xwb,Gayathri:2019kop,Yang:2020xyi,2021MNRAS.507.3362T,2021ApJ...915L..35K}). Some studies also shown a mixture of channels is strongly preferred over any single channel dominating the detected population (\citealt{2021ApJ...910..152Z}).

The latest population analysis carried out by LIGO-Virgo-KAGRA, GWTC3 (\citealt{LIGOScientific:2021djp,LIGOScientific:2021psn}), identified several population features that may be indicative of the binaries' origin. First, there appears to be a peak in the binary black hole mass spectrum around 30-40$\,M_{\odot}$ compared to a more simple power-law type population (\citealt{2021ApJ...913L..19T,2018ApJ...856..173T,2022ApJ...924..101E,2022PhRvD.105l3014S,Fishbach:2017zga}). At the same time a few observed black holes had unusual properties, such as masses in the so-called upper mass gap ($\gtrsim50\,M_{\odot}$), highly unequal masses in the binary, high spin and precessing mergers, which are rare in stellar evolution and might be indicative of alternative formation scenarios. 

Here we introduce a flexible approach to compare the predictions of detailed formation models with observations while simultaneously accounting for the potentially-confounding contributions from a flexible phenomenologically-parameterized model for compact binary formation.

Our paper is organised as follows.  In section 2 we introduce binary formation in AGN disks, phenomenological descriptions for formation in non-AGN sources,  and our flexible parametric population inference method. In section 3, we talk about the analysis's key findings. In section 4, we summarize our findings and comment on future directions. 

\begin{figure}[h!]
    \centering
	\includegraphics[scale=0.35]{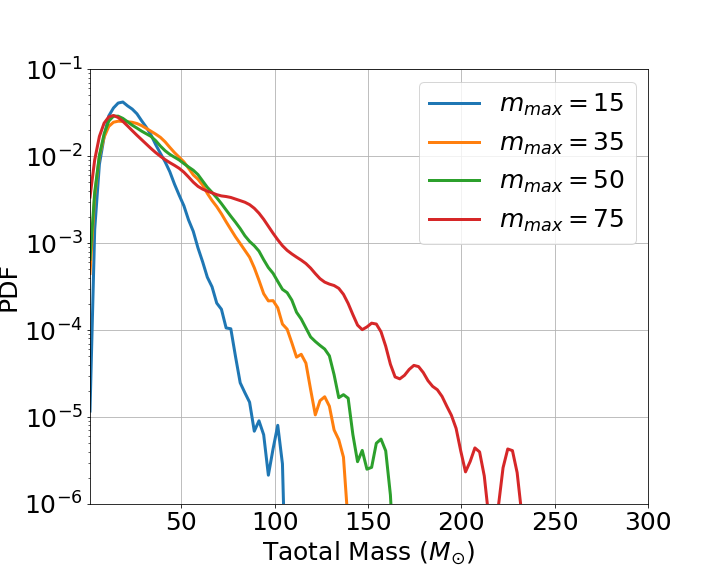}
	\includegraphics[scale=0.35]{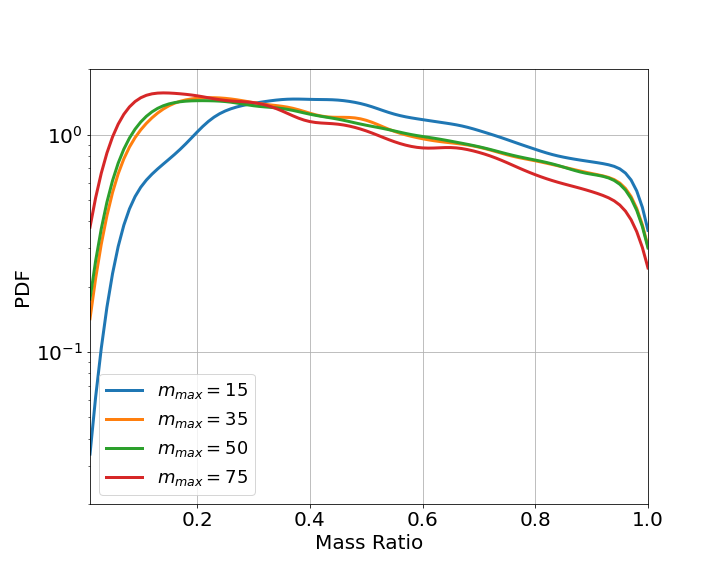}
	\caption{Parameter distributions for binary black holes formed in our AGN disk formation models; line color indicates the maximum natal BH mass.  As expected, as the maximum mass increases, total ($M$) mass upper limits increase. Additionally, a higher BH maximum natal mass $m_{\rm max}$ increases the relative frequency of  asymmetric mergers, particularly highly asymmetric mergers with $q<1/10$. }
	\label{fig:AGNModels:1d}
\end{figure}

\section{Methods}
\subsection{Binary mergers in AGN disks}
We construct a one-parameter model for BBH formation and merger within an AGN disk, parameterized by the maximum mass $m_{\rm max}$ of the natal BH distribution.
Specifically, we adopt a seed BH mass distribution which follows  the Salpeter mass function with index 2.35,  $dN/dm \propto m^{-2.35}$ with given $m_{max}$. For neutron stars (NS) we assume a normal distribution $m/M_{\odot} \sim N(1.49, 0.19)$. 
The BHs and NSs are assumed to orbit a supermassive black hole in an AGN, migrating into the disk and inward from their natal locations.   Close to the AGN, these objects undergo multiple encounters, facilitating binary formation and merger. Other AGN parameters are fiducial values that are expected to be typical, while there are large uncertainties. The range of possible values in the AGN models parameter space is discussed in (\citealt{McKernan_2018}).

Following (\citealt{2017ApJ...835..165B}), we adopted a geometrically thin, optically thick, radioactively efficient, steady-state accretion disk expected in AGNs. We used a viscosity parameter $\alpha = 0.1$, radioactive efficiency $\epsilon = 0.1$, fiducial supermassive BH mass $M_{\bullet}=10^6$\,M$_\odot$ and accretrion rate $0.1\dot{M}_{\rm Edd}$, where $\dot{M}_{\rm Edd}$ is the Eddington accretion rate.
Using \cite{2020ApJ...901L..34Y} and \cite{2020ApJ...898...25T,2020ApJ...899...26T}, we have computed the expected mass and spin distributions of binary mergers in AGNs.

Figure \ref{fig:AGNModels:1d} shows the binary black hole merger intrinsic parameter distribution for the AGN model with different initial mass limits ($m_{max}=[15M_{\odot},35M_{\odot},50M_{\odot},75M_{\odot}]$). As the natal BH mass upper limit $m_{max}$ increases, more massive binary components and total masses are allowed.  At the same time, for high $m_{max}$, asymmetric systems are more frequent.  By contrast, we have observed that the spin distribution properties are largely independent of our choice for $m_{max}$ limit.     

\subsection{Phenomenology of AGN and non-AGN sources}
\begin{figure}[h!]
    \centering
	\includegraphics[scale=0.3]{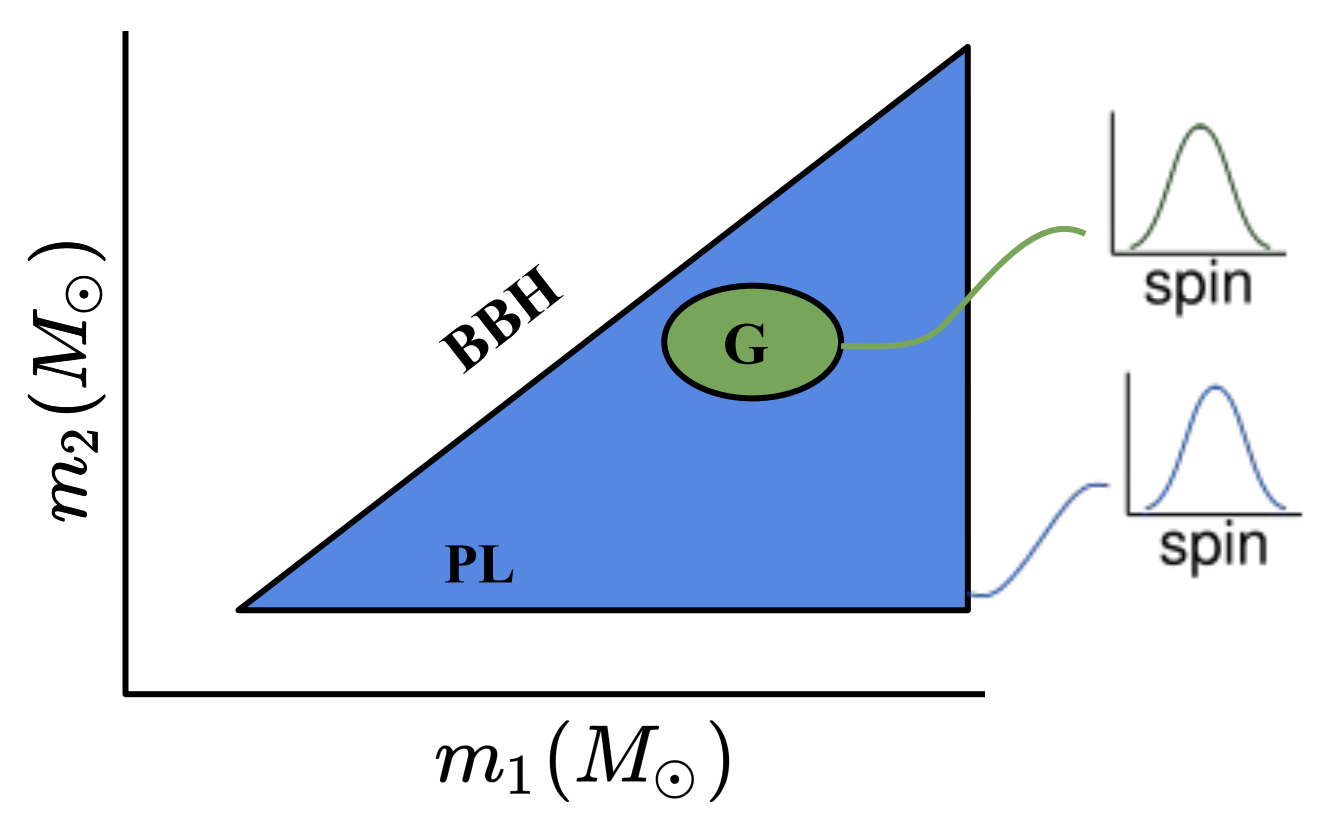}
	\includegraphics[scale=0.3]{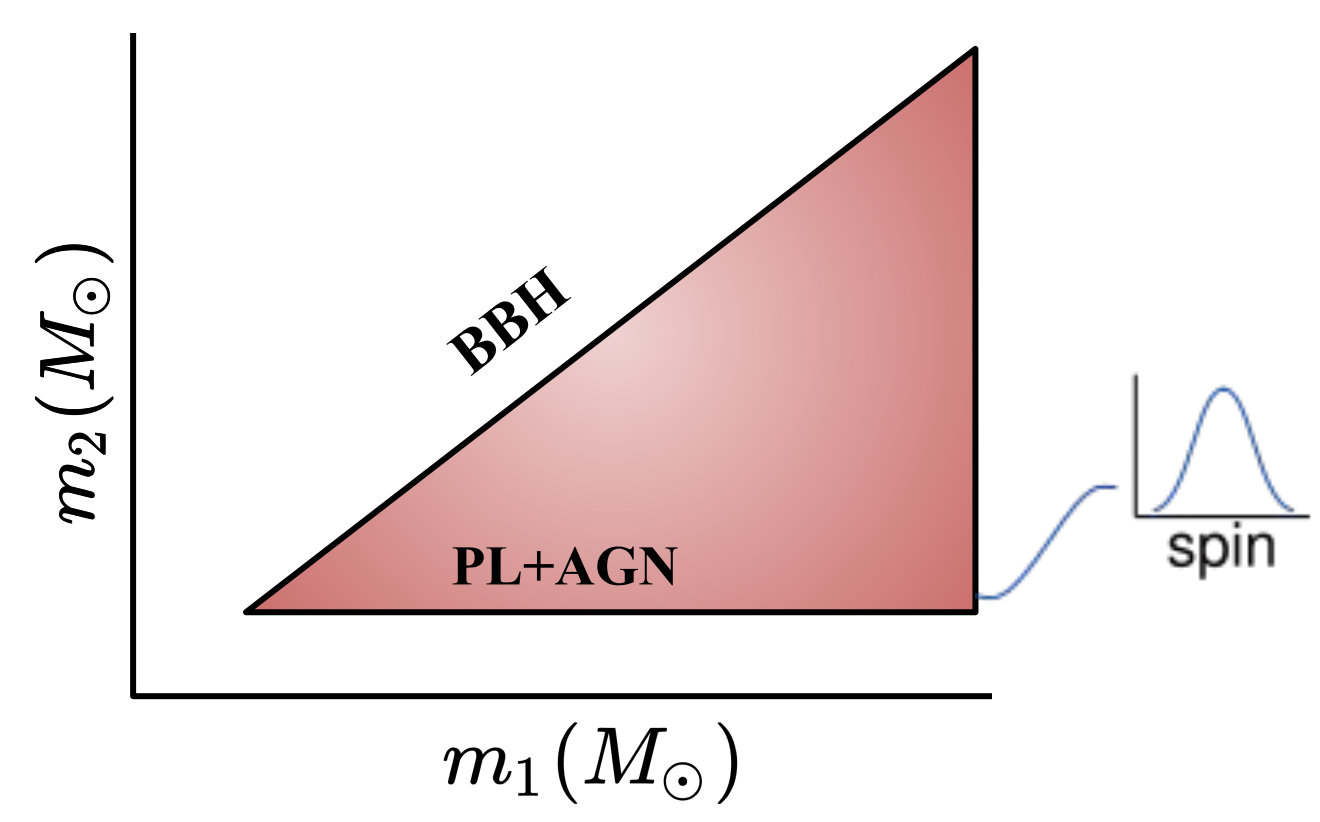}
	\includegraphics[scale=0.3]{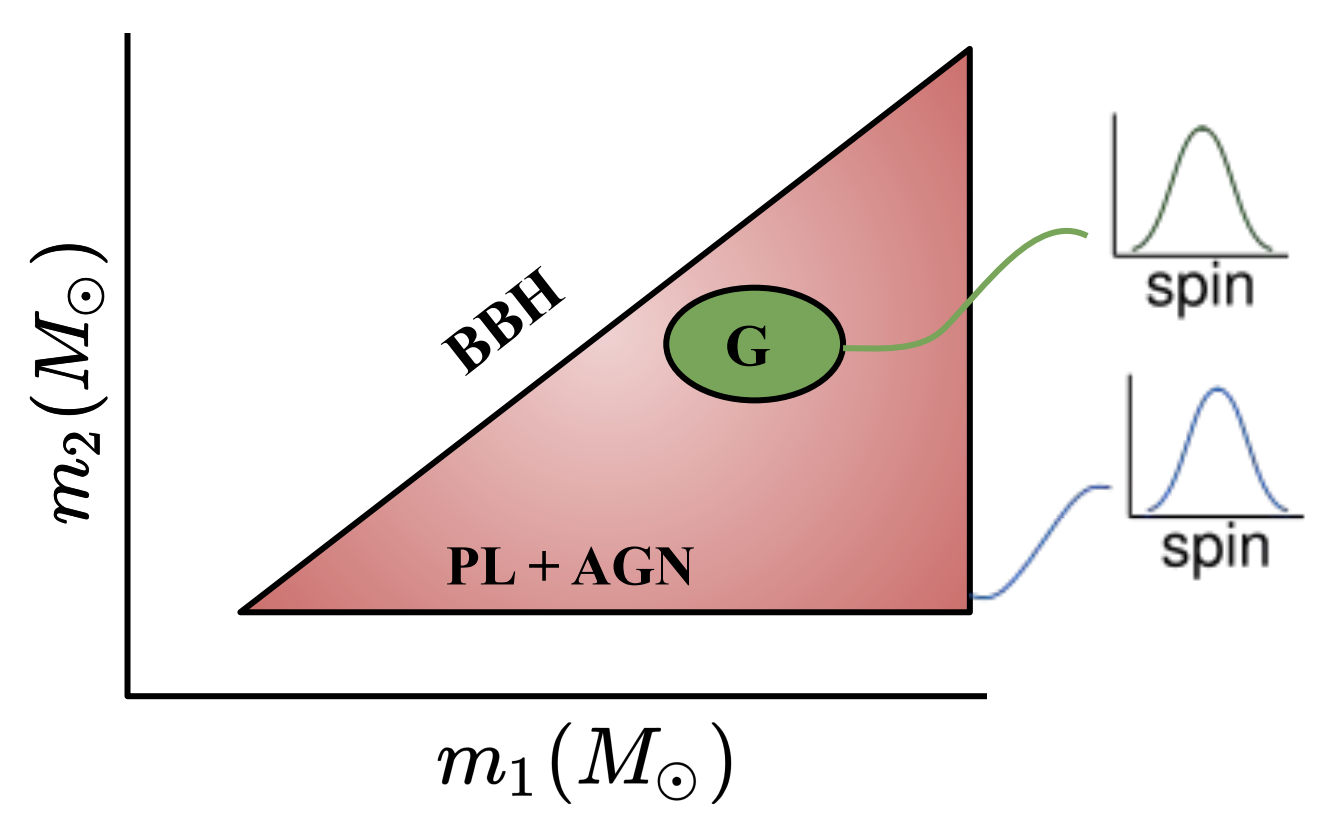}
	\caption{Graphical representations of the BBH population analysis in $m_1-m_2-$spin parameter. Top panel for the PL+G, the middle panel for the PL+AGN and the bottom panel for the PL+G+AGN model. }
	\label{fig:mixture_model}
\end{figure}

To allow for binary black holes which have a non-AGN origin, we follow previous work and introduce a few-parameter mixture model family.  As illustrated in Figure \ref{fig:mixture_model}, for the non-AGN component, we allow binary black holes to arise from a mixture of a power law and Gaussian components,  as detailed in Table 11 of \cite{2020arXiv201014533T}.   In the power law component, the primary is drawn from a pure power law mass distribution (with some unknown $m_{max,pl}<50 M_\odot$ and unknown primary power law index); the mass ratio is drawn from another power law; and the spins are drawn from an unknown Beta distribution.    In the Gaussian component, the primary and secondary are drawn from two independent Gaussian distributions with unknown mean and variance, with both means confined a priori to be near $30-40 M_\odot$ to be consistent with expectations for PISN supernovae. 
By including a second component this phenomenological model can allow for non-power-law features and also allow for spin distributions that vary with mass.
Each component $k$ has some undetermined overall rate ${\cal R}_k$.  The top panel of Figure \ref{fig:mixture_model} shows the general non-AGN model.

Our overall model is therefore a mixture model, parameterized by the unknown (continuous) AGN merger rate and its (discrete) maximum mass $m_{\rm max}$, along with all parameters of the non-AGN mixture model.  The overall merger rate density $dN/dVdXdt$ can therefore be expressed as a sum
\[
\frac{dN}{dVdXdt}= 
{\cal R}_{\rm agn} p_{agn}(X|m_{max}) 
+ {\cal R}_{g}p_g(X|\Lambda_g)
+ {\cal R}_{pl}p_{pl}(X|\Lambda_{pl})
\]
where $X$ are binary parameters and where $p_q,\Lambda_q$ are the model distributions and parameters for the $q$th component (AGN, Gaussian, and power-law respectively).

To systematically assess how well the distinctive features in AGN formation scenarios can be disentangled from this large model family, we will perform a sequence of calculations with increasing model complexity, as shown in the panels of Figure \ref{fig:mixture_model}.  Specifically, we consider without any non-AGN component; the power-law and Gaussian  (PL+G) model only; the power-law and AGN model (PL+AGN), without any Gaussian component; and finally the most general model with all three components.

\subsection{Population inference }
We describe and demonstrate a flexible parametric
method to infer the event rate as a function of compact binary parameters, accounting for Poisson
error and selection biases. In (\citealt{LIGOScientific:2020kqk}), analysed the Multi-Spin model which is the joint mass-spin model for binary black holes. Independent analyses have shown that there is a feature in the BBH mass spectrum around 30-40$\,M_{\odot}$, which is modeled as a Gaussian peak on top of a power law continuum. It is an empirical way of modeling extra features, but here we tried to understand it feature with AGN models. 

In this section we review the population inference \cite{Wysocki:2018mpo} used for multi-formation channel contributions. Binaries with intrinsic parameters
$x$ would merge at a rate $dN/dV_c \, dtdx = {\cal{R}}\,p(x)$, where $N$ is the number of detections, $V_c$ is the comoving volume, ${\cal{R}}$ is the space-time-independent rate of binary coalescence per unit comoving volume and $p(x)$ is the probability of $x$ from detected binaries.  The binaries intrinsic parameters includes mass $m_i$ and spins ${\bf{S}}_i$, where $i={1,2}$. 
The likelihood of the astrophysical BBH population at a given merger rate $\cal{R}$ (\citealt{2004AIPC..735..195L,2019MNRAS.486.1086M,2019PASA...36...10T}) and given binary intrinsic parameters $X\equiv(m_1,m_2,{\mathbf{\chi}}_1,{\mathbf{\chi}}_2)$, where $\mathbf{\chi}_i={\bf{S}}_i/m_i^2$, given the data for $N$ detections ${\cal{D}}=(d_1,...,d_N)$. This
the likelihood is given by

\begin{equation}
    {\cal{L}}({\cal{R}},X) \equiv p({\cal{D}} | {\cal{R}},X),
\end{equation}

\begin{equation}
  {\cal{L}}({\cal{R}},X) \propto e^{-\mu({\cal{R}},X)} \, \prod_{n=1}^N \int dx \,\ell_n(x) \,{\cal{R}}\, p(x,X),
\end{equation}
where $\mu({\cal{R}},X)$ is the expected number of detections under a given population parametrization $X$.  Using Bayes’ theorem one may obtain a posterior distribution on ${\cal{R}}$ and $X$ after assuming some prior $p({\cal{R}},X)$.   For computational efficiency and to enable direct comparison with discrete formation models, we use the Gaussian likelihood approximation technique introduced in (\citealt{2022arXiv220514154D}) 
to characterize each BBH observation's likelihood $\ell_n(x)$.

Using this formalism, we estimate what fraction of binary black holes are generated via the AGN channel using detected gravitational wave merger information. To do this study we have upgraded current parametric methods with a mixture model feature. Here we have a freedom to do the analysis with number of models which has different astrophysical binary distributions.

\section{Analysis}

As we discussed before, we perform population inference analysis with mixture model feature and detected confident detection binary black hole detection. For this study we have considered different the astrophysical binary distributions from AGN ( with different $m_{max}$), power-law, Gaussian peak and its combinations (see Figure \ref{fig:mixture_model}).

\subsection{Astrophysical merger rate}

\begin{table*}[t]
  \centering
  \begin{tabular}{|c|c|c|c|c|c|}
  \hline
    \bf{Analysis} & \bf{models}   & $m_{max}=15$ &$m_{max}=35$  & $m_{max}=50$ & $m_{max}=75$   \\ 
    \hline\hline
    PL only & $71.9^{+19.9}_{-20.4}$ & - & - & - & -  \\
    \hline
    G only & $19.1^{+2.5}_{-2.3}$ & - & - & - & - \\
    \hline
    AGN only &  & $84.7^{+19.5}_{-18.5}$  & $49.8^{+6.1}_{-5.5}$ & $52.9^{+7.8}_{-6.1}$ & $53.2^{+7.7}_{-6.2}$\\
       \hline
    PL+AGN  & PL & $29.3^{+9.9}_{-7.2}$  & $28.8^{+11.3}_{-6.8}$ & $25.7^{+7.6}_{-6.3}$ & $26.8^{+9.5}_{-6.2}$ \\
     \cline{2-6}
     & AGN &   $8.7^{+6.3}_{-4.5}$ & $8.3^{+8.3}_{-3.7}$ &  $12.7^{+6.5}_{-4.1}$&  $11.9^{+4.6}_{-3.2}$\\
  \hline
    PL+AGN+G & PL & $21.3^{+7.3}_{-5.2}$ & $23.5^{+6.6}_{-5.5}$ & $21.6^{+6.7}_{-5.0}$ & $23.3^{+7.1}_{-5.0}$ \\ \cline{2-6}
             & AGN & $6.7^{+5.6}_{-4.0}$ & $6.2^{+5.0}_{-3.5}$ & $12.7^{+5.0}_{-4.9}$ & $12.9^{+4.3}_{-4.3}$ \\ \cline{2-6}
             & G & $10.2^{+2.2}_{-2.3}$ & $9.3^{+2.6}_{-3.7}$ & $5.9^{+2.9}_{-1.9}$ &$5.6^{+2.2}_{-1.5}$ \\ \hline
    
  \end{tabular}
  \caption{The astrophysical rates from PL+AGN and PL+AGN+G models. Each row corresponds to each analysis and each column corresponds to different AGN models with different initial mass limit. }
  \label{tab:1}
\end{table*}

We have estimated the astrophysical black hole merger rate for a given astrophysical model with a given number of GW detection. For this study, we have used the parameter estimation samples from obtained by LIGO-Virgo-KAGRA Collaboration  (\citealt{LIGOScientific:2021usb,LIGOScientific:2021qlt,2018arXiv181112907T,LIGOScientific:2020ibl}) and it available in the Gravitational Wave Open Science Center (https:
//www.gw-openscience.org) with the mixed model. Here we follow the same selection criteria as (\cite{LIGOScientific:2021psn}), we have considered events with a false alarm rate of $< 0.25 yr^{-1}$. 

Table  \ref{tab:1}  shows the merger rates inferred for joint  PL only, G only, AGN only (with different $m_{max}$), PL+AGN model (with different $m_{max}$) and PL+G+AGN models (with different $m_{max}$). 
 We have estimated the merger rate results derived using each individual model component as well as combined. We have observed that the inferred merger rate from the AGN-only models with different choices for $m_{\rm max}$ produces largely consistent results peaking near $50  \,Gpc^{-3} yr^{-1} $.  For the smallest maximum BH natal mass  $m_{max}=15 M_\odot$, the inferred single-component AGN merger rate peaks around $80 \,Gpc^{-3} yr^{-1}$. For all other models, they peak around the same $\cal{R}$.  
Similarly, we have inferred the merger rate distribution for the single-component Gaussian, and power-law components.  As  expected, the merger rate from the power-law component dominates overall, as it incorporates and describes many frequent mergers of the lowest-mass binary black holes.
In the case of PL+AGN, the inferred AGN and PL components have a well-determined merger rate, with median AGN merger rate $\simeq 8/, {\rm Gpc}^{-3} {\rm yr}^{-1}$ and PL merger rate $\simeq 30/, {\rm Gpc}^{-3} {\rm yr}^{-1}$.
Changing the maximum natal BH mass  $m_{max}$ has a very mild impact on the inferred AGN merger rate, and almost no effect on the inferred PL merger rate.

In the case of PL+G+AGN analyses, the inferred AGN, G and PL components have a well-determined merger rate, with median AGN merger rate $\simeq 7/, {\rm Gpc}^{-3} {\rm yr}^{-1}$, G merger rate $\simeq 4/, {\rm Gpc}^{-3} {\rm yr}^{-1}$ and PL merger rate $\simeq 30/, {\rm Gpc}^{-3} {\rm yr}^{-1}$. As we have seen in the PL+AGN study, we have not seen any major effect on PL or G merger rate when we change the AGN model.
Our inferred merger rates deduced from the multi-component model are consistent with inferences performed using single-component models alone, suggesting that inference isolates only the contribution from each component.

\subsection{Inferred merger rate versus mass}
\begin{figure}

      \hspace*{-1cm}
         \includegraphics[scale=0.25]{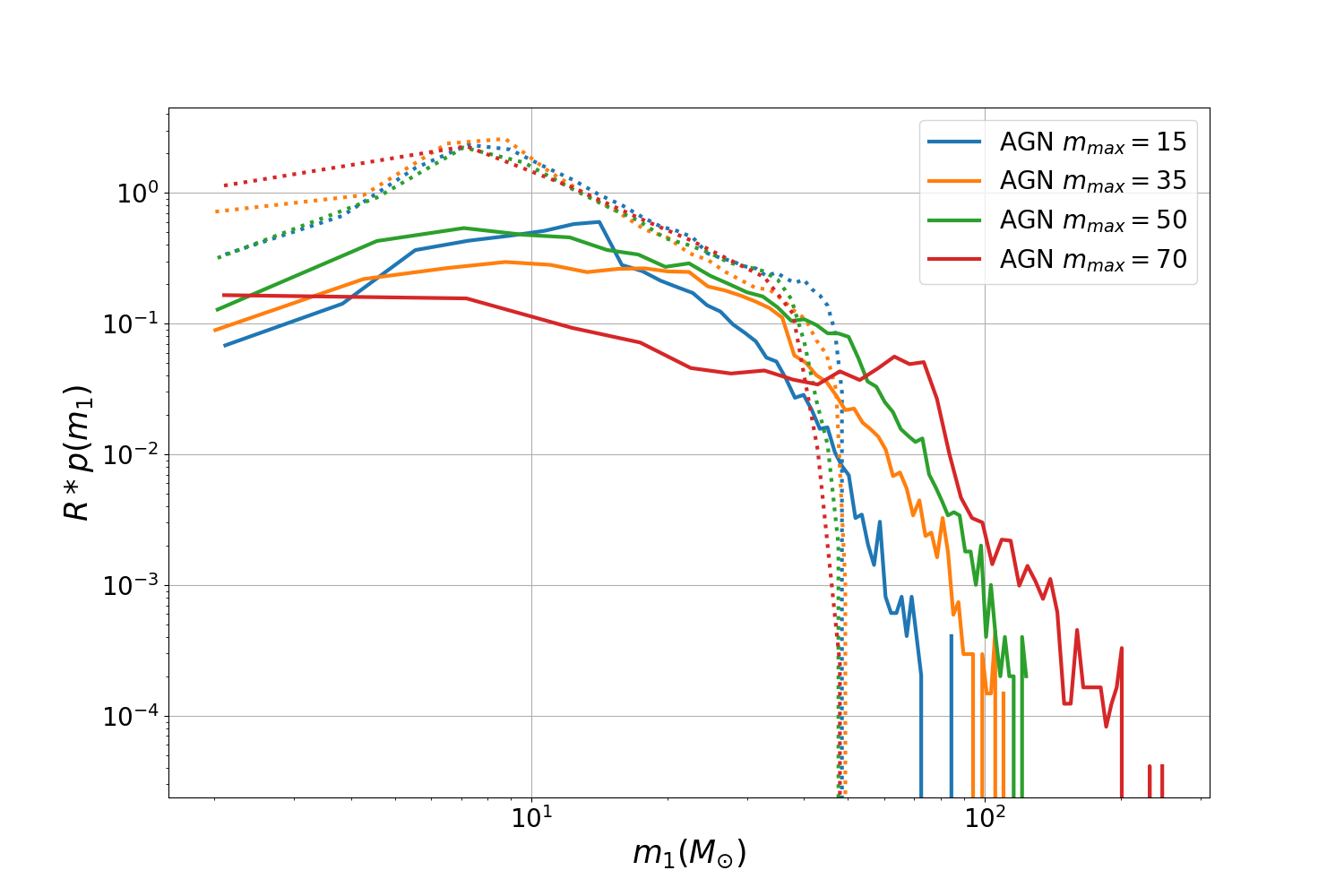}
         \label{fig:ppd_plagn}
     \hfill
    
         \hspace*{-1cm}
         \includegraphics[scale=0.25]{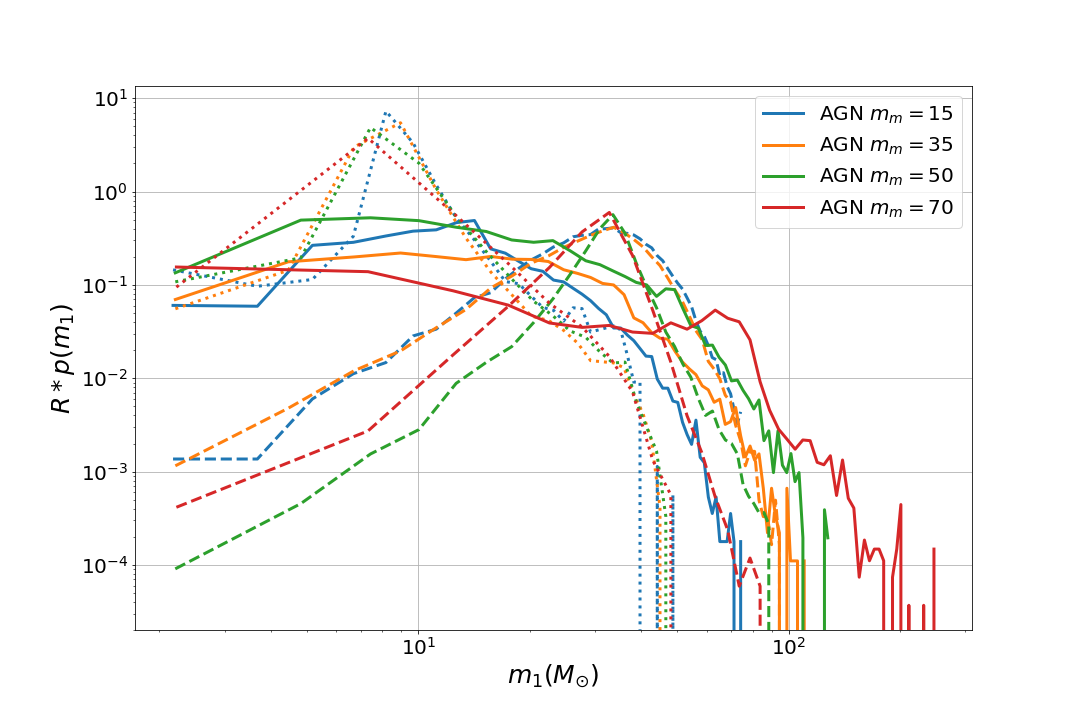}
         \label{fig:ppd_plgagn}
\caption{The inferred merger rate versus mass for PL+G+AGN and PL+G+ AGN model analyses. The solid, dashed and dotted lines for AGN, G and PL models components.}
\end{figure}

To better appreciate how well our inference directly projects out the relative contribution from each component, Figure \ref{fig:ppd_plagn} and \ref{fig:ppd_plgagn} shows our inferred merger rate versus mass for PL+AGN and PL+G+AGN models analyses respectively. In each plot we have shown each model component in mass space. 

As we expected the low mass region is highly contributed by the PL model compared to other models. We have observed a peak in PL model distribution around $7 M_{\odot}$ for both PL+AGN as well as PL+G+AGN analyses. The peak is prominent for  the PL+G+AGN study compared to PL+AGN. The high mass region is represented by only AGN, as we expected to see. Note that, the AGN model not only contributes in high mass region it also contributes full mass space as shown in \ref{fig:ppd_plagn} and \ref{fig:ppd_plgagn}. 

\subsection{Inferred merger rate versus mass ratio}

Similarly here we show the inferred merger rate versus mass ratio for PL+AGN and PL+G+AGN models analyses. 

\begin{figure}

      \hspace*{-1cm}
         \includegraphics[scale=0.25]{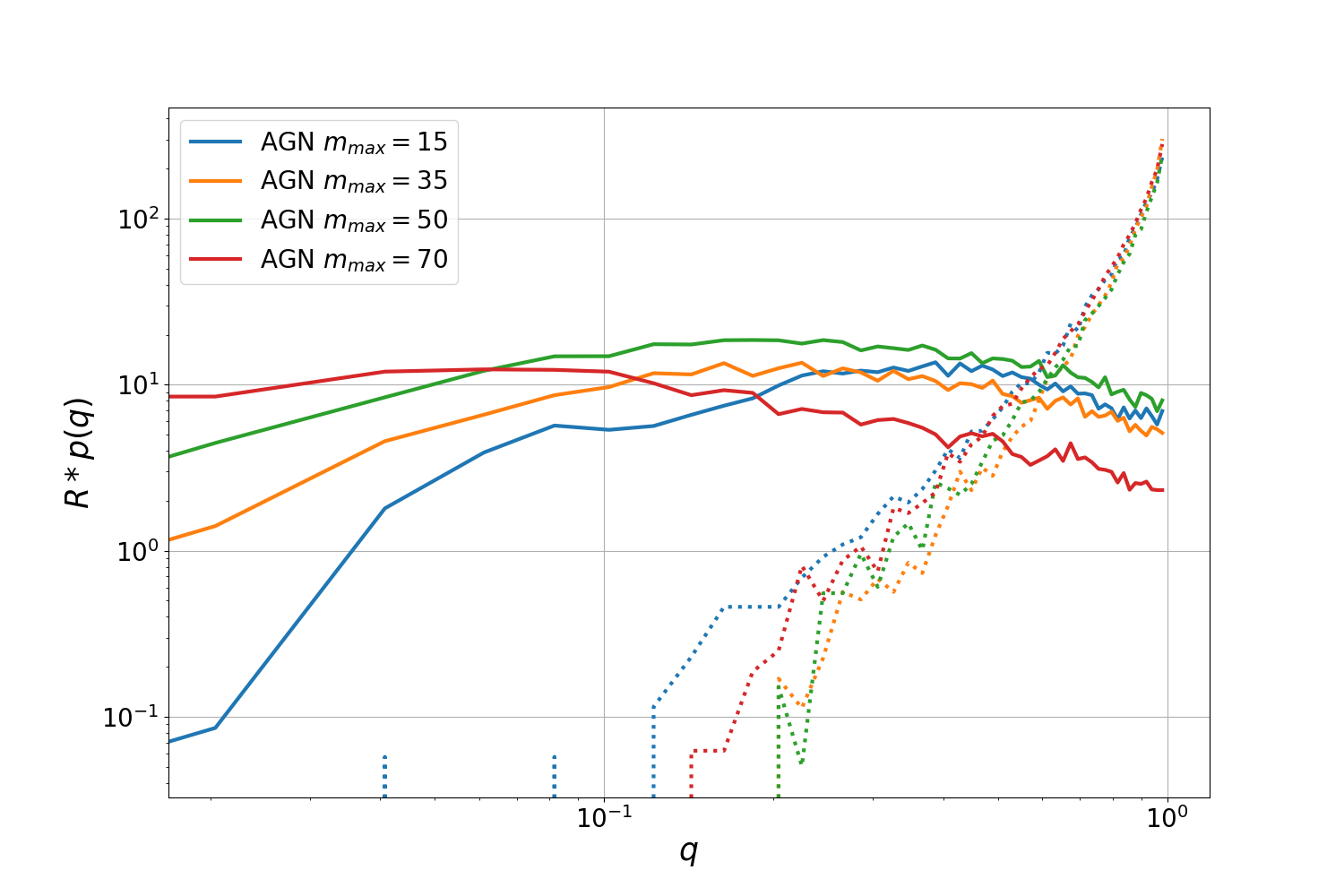}
         \label{fig:ppd_plagn2}
     \hfill
    
         \hspace*{-1cm}
         \includegraphics[scale=0.25]{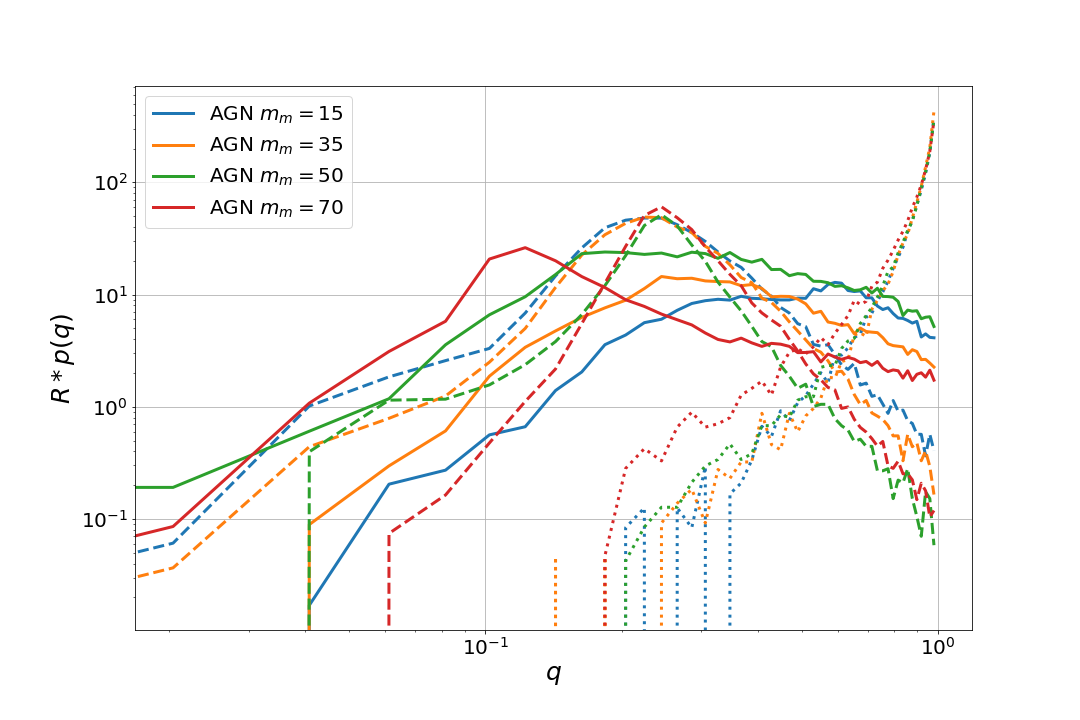}
         \label{fig:ppd_plgagn2}
\caption{The inferred merger rate versus mass ratio for PL+AGN and PL+G+ AGN model analyses. The solid, dashed and dotted lines for AGN, G and PL models components.}
\end{figure}
Figure \ref{fig:ppd_plagn} and \ref{fig:ppd_plgagn} shows our inferred merger rate versus mass ratio for PL+AGN and PL+G+AGN models analyses respectively. In each plot we have shown each model component in mass ratio space. As we expected the low mass ratio region contributed by AGN model for PL+AGN analysis and AGN \& G models for PL+G+AGN analysis. For high $q$, the dominate contribution from  PL model, that is consistence with detected events.

\subsection{Power-law model parameters }

As we discussed before, the contribution of a power-law model to the overall merger rate does not change substantially if we include or omit other model components like AGN or G.  Among the models we consider, this quasi-universality is expected: the PL model most effectively reproduces the merger rate versus mass for the lowest-mass and most frequently merging binary black holes.  While the overall merger rate from this component is stable to our choice of the mixture, the model parameters recovered for PL depend strongly on which other confounding contributions are also present,
as suggested by Figure \ref{fig:ppd_plagn} and \ref{fig:ppd_plgagn}.
While the PL mass ratio distribution does not depend strongly on including or omitting AGN or G, the power law slope $\alpha$ and minimum mass $m_{min}$ do change substantially. The estimated $\alpha$ median value with 65\% credible intervals from different analysis as $1.6^{+0.2}_{-0.2}$, $6.7^{+3.1}_{-2.4}$,$8.4^{+2.3}_{-3.4}$, and $1.8^{+0.5}_{-0.5}$ for PL-only, PL+G, PL+AGN ($m_min$=50) and PL+G+AGN ($m_min$=50) respectively. Similarly, $m_{min}$ estimation are $2.5^{+0.3}_{-0.3}$, $8.4^{+0.2}_{-0.3}$, $8.5^{+0.2}_{-0.4}$, and $6.7^{+1.6}_{-1.3}$ for PL-only, PL+G, PL+AGN ($m_min$=50) and PL+G+AGN ($m_min$=50) respectively. 
For example, the $\alpha$ estimation suggests that while a pure power-law model favours  $m_{\rm min}$ close to the lower limit our priors allow,  incorporating other components causes the power-law component's minimum mass to favour larger masses.  With the pertinent mass range for the power-law changing substantially via different $m_{\rm min}$, unsurprisingly.  The $\alpha$ estimation has a wide range of inferred power law exponents, as the PL may dominate only an extremely narrow range of masses; see Figure \ref{fig:ppd_plagn}.

\section{Conclusion}

In this paper, we have directly compared a one-parameter model for AGN binary black hole formation with the reconstructed sample of binary black holes identified via gravitational wave observations.  
To deconvolve the AGN component from binaries with different origin, we allow for BBH formation in both AGN and phenomenological channels.  We consistently find a significant contribution to the merger rate from the AGN component ($\simeq O(5/{\rm Gpc}^{-3}{\rm yr}^{-1}$).  Our inferred AGN contribution follows by our prior belief on the maximum mass of BBH formed from other channels, which we presume is less than $50 M_\odot$ due to pair-instability impacts on stellar evolution and death.

As in previous studies (\citealt{Yang_2020,Yang:2020xyi,Gayathri:2019kop,Gayathri:2021xwb,Vajpeyi_2022}), our models for AGN BBH formation predict a wide range of BBH mass ratios and frequent significant spins.  At present, because the distinctive signatures of AGN formation are preferentially imparted only to the most massive BBH, the extant BBH sample does not yet contain enough events to provide overwhelming evidence in favour of an AGN component, consistent with prior work (\citealt{Vajpeyi_2022}) 
subsequent observations could support or rule out this channel.

\noindent{\bf Acknowledgements}
We gratefully acknowledge the support of LIGO and Virgo for the provision of computational resources. G.V. and D.W. acknowledge the support of the National Science Foundation under grant PHY-2207728.  I.B. acknowledges the support of the National Science Foundation under grants \#1911796, \#2110060 and \#2207661 and of the Alfred P. Sloan Foundation. This research has made
use of data, software and/or web tools obtained from
the Gravitational Wave Open Science Center (https:
//www.gw-openscience.org), a service of LIGO Laboratory, the LIGO Scientific Collaboration and the Virgo
Collaboration.
LIGO is funded by the U.S. National Science Foundation. Virgo is funded by the French Centre National de Recherche Scientifique (CNRS), the Italian Istituto Nazionale della Fisica Nucleare (INFN) and the
Dutch Nikhef, with contributions by Polish and Hungarian institutes. This material is based upon work supported by NSF's LIGO Laboratory, which is a major facility fully funded by the National Science Foundation.

\bibliography{reference}

\end{document}